\documentclass[12pt]{iopart}
\usepackage{graphicx}  
\begin{document}

\title[Point Contact Spectroscopy measurements of Cobalt doped Ba-122]{$a-b$ Plane Point Contact Spectroscopy measurements of optimally Cobalt doped Ba-122 iron-pnictide superconductors}

\author{J Timmerwilke, E Kim, J Maughan, JS Kim, GR Stewart and A Biswas}

\address{Department of Physics, University of Florida, Gainesville, FL 32611}
\ead{\mailto{timmerwi@phys.ufl.edu},\mailto{amlan@phys.ufl.edu}}
\begin{abstract}
Point contact spectroscopy (PCS) is a technique which can reveal the size and symmetry of a superconducting gap ($\Delta$) and is especially useful for new materials such as the iron-based superconductors. PCS is usually employed in conjunction with the extended Blonder-Tinkham-Klapwijk (BTK) model which is used to extract information such as the existence of nodes in $\Delta$ from PCS data obtained on unconventional superconductors. The BTK model uses a dimensionless parameter $Z$ to quantify the barrier strength across a normal metal - superconductor junction. We have used a unique feature of PCS which allows variation of $Z$ to obtain crucial information about $\Delta$. We report our $Z$-dependent, $a-b$ plane PCS measurements on single crystals of the iron-based superconductor BaFe$_{2-x}$Co$_{x}$As$_{2}$. Our measurements show that BaFe$_{2-x}$Co$_{x}$As$_{2}$ ($x$=0.148) is a superconductor with two gaps which does not contain any nodes. The $Z$ dependent point contact spectra rule out a pure $d$ symmetry and the gaps at optimal doping have negligible anisotropy.
\end{abstract}

\maketitle

\section{Introduction}
\subsection{The superconducting gap}
In studying superconductors it is generally accepted that knowing the size and symmetry of superconducting gap(s) is key to recognizing and understanding the underlying cause of the superconductivity.  For example, in superconductors such as MgB$_2$, evidence of multiple gaps associated with different branches of the Fermi surface led to a better understanding of the mechanism responsible for the high $T_c$s observed in the binary compound ~\cite{choi2002}. A similar statement can be made about the cuprates, where the symmetry of the gap clearly shows that the behavior cannot be explained by the Bardeen-Cooper-Schrieffer (BCS) model ~\cite{0034-4885-62-1-002}. Point Contact Spectroscopy (PCS) is capable of providing extensive information about the SG and has proved itself to be a robust method especially for dealing with new superconductors, such as the iron pnictides ~\cite{PhysRevLett.105.237002,Chen2008,RefWorks:28,RefWorks:1,RefWorks:29}.  PCS has less stringent requirements for surface quality compared to scanning tunnelling spectroscopy and is mechanically stable since the two electrodes are in direct physical contact.

\subsection{Probing gap symmetry using point contact spectroscopy}
In PCS, spectroscopic information can be extracted by varying the voltage across a small, ballistic or near ballistic, junction between two metals and measuring the differential conductance ($dI/dV$) as a function of $V$~\cite{Fisher:172}.  
In the case that one of the junction electrodes is replaced with a superconductor, the Blonder, Tinkham, and Klapwijk (BTK) model can explain the spectra for simple $s$-wave superconductors ~\cite{RefWorks:7}.  The BTK model assumes a one dimensional junction between a superconductor and normal metal with the barrier between them represented by the function $U(x)=H\delta(x)$. The current depends on three quantities {\em viz}., the temperature, the gap size, and a dimensionless constant $Z =H/\hbar v_{F}$, where $v_F$ is the Fermi velocity.  The shape of the differential conductance depends strongly on $Z$ for energies less than the gap.  For small values of $Z$ (high transparency junctions), Andreev reflection dominates and a doubling of the conductance occurs for $E < \Delta$. For large values of $Z$ (low transparency junctions) where the contact behaves like a tunnel junction, the conductance is proportional to the local density of states~\cite{PhysRevLett.5.147}. $Z$ can be varied in a point-contact junction unlike a standard tunnel junction, which is an additional advantage of PCS since the variation of the point contact spectra with $Z$ is characteristic of particular gap symmetries.

The BTK model needs to be extended to account for compounds which are not pure BCS superconductors.  We focused on three specific extensions {\em viz.}, broadening effects from the finite lifetime of the quasiparticles, the possibility of more than one gap, and anisotropic gaps and gap symmetries which contain nodes or sign flips.  A beneficial review of extended BTK models and how to implement them can be found in ~\cite{RefWorks:5}. The broadening effect results in a smearing of features, which is superficially the same as an increase in temperature. This effect is included by the addition of an imaginary energy term $i\Gamma$. This term represents the finite lifetimes of the electron-like and hole-like quasiparticles.  $\Gamma$ is often larger than what would be predicted from just the lifetimes because it accounts for many inelastic scattering events which can occur at the junction of the two electrodes. Hence, reducing $\Gamma$ is an important aspect of accurate PCS.

Adding multiple gaps to the BTK model consists of calculating a conductance for each individual gap, multiplying it by a weight factor, and summing the conductance due to each corrected gap.  In the case of a two gap system only a single weight factor ($w$) is needed and the total conductance is expressed as $G(V)=w G_1(V)+ (1-w)G_2(V)$.  In such a measurement each gap can have separate $Z$ and $\Gamma$ values.  Thus the number of variables necessary for a fit quickly becomes large and it is increasingly difficult to obtain accurate fits as the number of gaps increases.

Important to our experiment is the ability to differentiate between different gap symmetries and identify the presence or lack of a node. Kashiwaya {\em et al.} extended the BTK model to non-BCS superconductors for current injection into the $a-b$ plane~\cite{RefWorks:23}.  The shape of conductance in this extended model measured at various $Z$ values differ dramatically.  For large $Z$, when a sign flip is present (e.g. for $d$-wave symmetry), a strong zero-bias conductance peak is predicted by theory and has been observed through tunnelling experiments in cuprates~\cite{Aprili20001864}.  The lack of nodes leads to a conductance which flattens inside the gap region and if nodes exist then the conductance forms a peak like feature at zero bias.  Figure 1 shows the various $dI/dV-V$ curves expected for gaps with no nodes, with nodes, and a sign flip in the symmetry, in both the low $Z$ and high $Z$ conditions calculated using refs.~\cite{RefWorks:5,RefWorks:23}.  Comparison of the shape of our measured conductance to the modeled data will allow us to identify the symmetry of the superconducting gap.  Our full model includes two gaps ($\Delta_1$ and $\Delta_2$), two $Z$ values ($Z_1$ and $Z_2$), two $\Gamma$ values ($\Gamma_1$ and $\Gamma_2$), a weight factor ($w$), and an angular dependence of the gaps.

\subsection{Gap symmetry of Iron-based superconductors}
Among the multiple families of the iron-based superconductors (FeSCs), we chose the AFe$_{2}$As$_{2}$ (122) compounds (A represents ions such as Ca, Ba, and Sr) since they can be grown as single crystals of approximate size $1$mm $\times$ $1$mm $\times$ $0.1$mm and larger~\cite{PhysRevLett.101.117004},  which aids PCS measurements.  The 122 compounds can also incorporate an extensive range of elements which can be used as dopants for isovalent, electron, and hole doping, leading to rich and informative phase diagrams.  The comparison between these different dopings can hopefully be used to identify important underlying features related to the superconductivity in FeSCs and perhaps other compounds~\cite{RevModPhys.83.1589,0295-5075-90-3-37005}.
We focused on BaFe$_{2-x}$Co$_{x}$As$_{2}$ (Co-doped Ba-122) compounds. Significant research has already been performed on these cobalt doped compounds, using both bulk techniques such as specific heat and surface techniques such as PCS to examine the gap directly and indirectly. To put our results in context, we list the salient features of previous research on Co-doped Ba-122. Gofryk {\em et al.}~\cite{PhysRevB.81.184518} obtained evidence that the gap structure cannot be described by a single isotropic gap from specific heat measurements, requiring instead an anisotropic structure or multiple gaps.  They observed a negligible (less than $10\%$) variation in the ratio of the small gap to $T_c$ with doping ($x$ = 0.09, 0.16, 0.206, and 0.21).  A $30\%$ reduction in the ratio was observed for the large gap in underdoped samples.  Luan {\em et al.}  took penetration depth measurements and obtained evidence for two full gaps in the underdoped region ($x$ = 0.1), with possible signs of a coexistent magnetic phase ~\cite{PhysRevB.81.100501}.  Gordon {\em et. al.} took penetration depth measurements over a large number of dopings, finding a sharp increase in the in-plane penetration depth in the underdoped region, which is consistent with the gap developing greater anisotropy as it goes to the superconducting dome's edge ~\cite{PhysRevB.82.054507}.  Reid {\em et al.} took heat transport measurements which showed evidence of nodes for both under and over-doped crystals, but which lessened as optimal doping was approached ~\cite{PhysRevB.82.064501}.  Two gaps of sizes 3.1 meV and 7 meV were also observed in ARPES data by van Heumen {\em et al.} ~\cite{0295-5075-90-3-37005} .  Point contact measurements taken by Tortello {\em et al.} ~\cite{PhysRevLett.105.237002} showed clear evidence for a two gap system that excludes a $d$-wave symmetry($x$ = 0.2).  Point contact measurements have also been taken by Samuely {\em et al.}~\cite{RefWorks:1} who observe a single gap ($x$=0.2) and Lu {\em et al.}~\cite{RefWorks:28} obtained a sharp conductance peak, but no gap ($x$ = 0.2).   
The superconductivity in FeSCs might be caused by spin fluctuation pairing, as described in the $s\pm$ model ~\cite{RefWorks:8}.  We cannot rule out such a model since for the values of $\Gamma$ in our measurements we cannot distinguish the ZBCP that might be seen in the $s\pm$ model~\cite{Nagai2010S504}.
The degree of anisotropy in this model is being debated and in certain cases may contain accidental nodes ~\cite{PhysRevB.83.220508}. It is reasonable given the available measurements that the two gaps as observed by Tortello {\em et al.} with little anisotropy for the optimal doping is most likely the correct interpretation. However, the PCS measurements mentioned above do not vary the $Z$ parameter and this aspect of PCS measurements is necessary to provide further test of the anisotropy.

In this paper, we report our $Z$-dependent, $a-b$ plane PCS measurements on BaFe$_{2-x}$Co$_{x}$As$_{2}$ ($x$ = 0.148). As described above, the $Z$-dependence of PCS can be used to distinguish between different gap symmetries and has not been performed systematically in the Co-doped Ba-122 system. In addition, the extended BTK models require that the current is injected into the $a-b$ plane of the superconductor, which we have achieved through a technique used earlier in $n$-doped cuprates ~\cite{PhysRevB.68.024502}. We find that our $Z$ dependent measurements are consistent with a two $s$-wave gaps with no evidence for nodes.

\section{Experimental Methods}

Our samples are optimally doped single crystals of BaFe$_{2-x}$Co$_{x}$As$_{2}$ ($x$ = 0.148), with a transition temperature ($T_c$) of 25 K and a transition width of $\sim$ 1 K as observed in four probe resistivity measurements.  The crystals were grown using the flux method with an Indium flux, similar to that used in ~\cite{0953-8984-21-34-342201}.  A ratio of Ba:(Fe,Co):As:In of 1:2:2:20 was heated from room temperature to 1100$^\circ$ C at 75$^\circ$ C/hr, then cooled to 600$^\circ$ C at 5$^\circ$ C/hr, before finally being returned to room temperature at 75$^\circ$ C/hr. To obtain $Z$-dependent PCS data along the $a-b$ plane, we required an apparatus capable of varying the point contact junction resistance while injecting the current parallel to the ab-plane of the crystal.  There are two methods of obtaining PCS data {\em viz.}, soft PCS in which the junction is formed using a small metal contact on the surface of a superconductor and hammer and anvil PCS in which a tip is brought into close contact with the superconductor, mechanically or with a piezoelectric ~\cite{RefWorks:5}.  Cobalt doped Ba-122 has been measured using the soft PCS method by Tortello {\em et al.}~\cite{PhysRevLett.105.237002}, but this method does not allow one to change the $Z$ parameter on a single sample.  To vary $Z$, we designed a mechanical device which lowers a thin tungsten wire onto the sample in a controlled manner (figure 1 inset), similar to the system used in ~\cite{PhysRevB.68.024502}. Our probe utilizes a mechanical cantilever to lower the tungsten wire horizontally onto the edge of the sample.  This method is different from the normal hammer and anvil PCS and it is possible that the contact area could be quite large resulting in diffusive junctions.  PCS measurements require that the junction be ballistic or nearly ballistic, such that the injected electrons do not go through inelastic collisions within the junction.  As the area of a junction increases the likelihood of inelastic collisions occurring increases resulting in large values of $\Gamma$ and a loss of spectroscopic information.  However, our measurements showed energy resolved spectroscopic data and thus we can conclude that, similar to the case of soft PCS where the contact area is also large, our junction is dominated by one or many smaller micro-junctions at the nanometer scale. The stability of our junctions at different temperatures was still an issue, as we desired junctions which were stable above $T_c$ down to our minimum temperature of 4.2 K.  Micro-junctions in an insulating material are known to be responsive to pulses of high voltage, which can either create or destroy new junctions ~\cite{PhysRevLett.105.237002}.  Hence, we decided to apply a voltage of about 100 mV during the process of moving the wire onto the sample.  We found this process increased the stability of the junction during the cooling process and thereafter.

\begin{figure}
  \includegraphics[width=8.5cm]{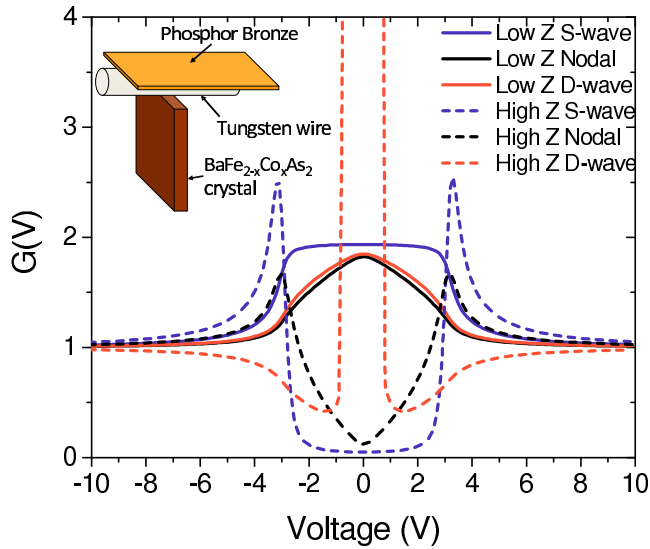}
  \caption{Modeled conductance curves for single gaps with three different symmetries each at two different $Z$ values.  The inset shows a schematic of the point contact setup used in our measurements.(Colour in online only)}
  \label{figure 1}
\end{figure}

	Joule heating is also a concern in PCS.  While it is difficult to completely rule out Joule heating, it can be discounted as a significant effect by looking at the derivative of the conductance curve outside of the range in which the superconducting gap dominates.  If the slope of the conductance curve is the same as the sign of the voltage, i.e. positive for positive voltages, then Joule heating is likely not a significant effect ~\cite{0953-8984-21-9-095701}.  Our data suggest that Joule heating is not significant for energies up to $2\Delta$. The apparent $T_c$ of a superconductor can also change due to Joule heating. Although we do observe lower $T_c$'s (the resistance transition occurs 6K higher than any features we can observe using PCS) than those measured using four probe resistance measurements in our samples, this reduction is possibly due to degradation of the surface layer of the superconductor since FeSCs have shown evidence of such degradation with exposure to air~\cite{RefWorks:28}.

\section{Results and Discussion}

Figure 2a shows $dI/dV-V$ data for low $Z$ measurements on a BaFe$_{2-x}$Co$_{x}$As$_{2}$ ($x$ = 0.148) crystal.  The high $Z$ measurement does not show the individual superconducting features, but rather a V-shaped conductance curve with a broad peak like feature between from 20 meV to about 40 meV.  These features are tied to the superconductivity since they disappear when the temperature is increased above the $T_c$ of the sample.  However, in repeat measurements where such features are seen, the energy at which these features appear varies in different junctions by roughly 20 meV (ranging between 20 meV out to 60 meV, though peaks near 20 meV were more common). The $Z$ was then reduced by pressing the tungsten wire further on the single crystal. The $dI/dV-V$ curves for the low $Z$ junctions have to be normalized to enable comparison to both the BTK model and between measurements at different temperatures.  The normalization method relies on the assumption that the conductance goes to a constant value at energies of about 2$\Delta$ and greater as suggested by the extended BTK model. We have normalized our data in two steps. First, the $dI/dV-V$ curve is divided by the value of $dI/dV$ at $V$= 25 meV, which is more than twice the voltage at which gap features present, to obtain $dI/dV_{norm1}$.
Figure 2b shows the $dI/dV_{norm1}-V$ data for a low $Z$ junction at 4.2 K and 21 K, in which we see evidence for a two-gapped system with approximate gap values of 10$\pm$1 meV and 6$\pm$1 meV as marked by the vertical lines.  These gaps can be compared to point contact measurements taken by Tortello {\em et al.} for a slightly over-doped Ba-122 sample, where they found gaps of 4.4 meV and 9.9 meV in the $a-b$ plane~\cite{PhysRevLett.105.237002}. To compare the data to the BTK model we should obtain a flat $dI/dV-V$ curve at temperatures above $T_c$.  There is a slight peak feature seen in the 4.2 K data near 0 meV, this is seen in a few other curves and will be discussed later. It is clear in figure 2b that at 21 K the superconducting features have been suppressed but the conductance is not flat. Hence, in the second step of our normalization procedure we remove the parabolic background observed in the curve at 21 K by dividing the $dI/dV_{norm1}-V$ data at all the temperatures by the $dI/dV_{norm1}-V$ curve at 21 K to obtain $dI/dV_{norm2}-V$ curves as shown in figure 3.  This normalization process facilitates accurate identification of the gap structure and size.

\begin{figure}
  \includegraphics[scale=0.75]{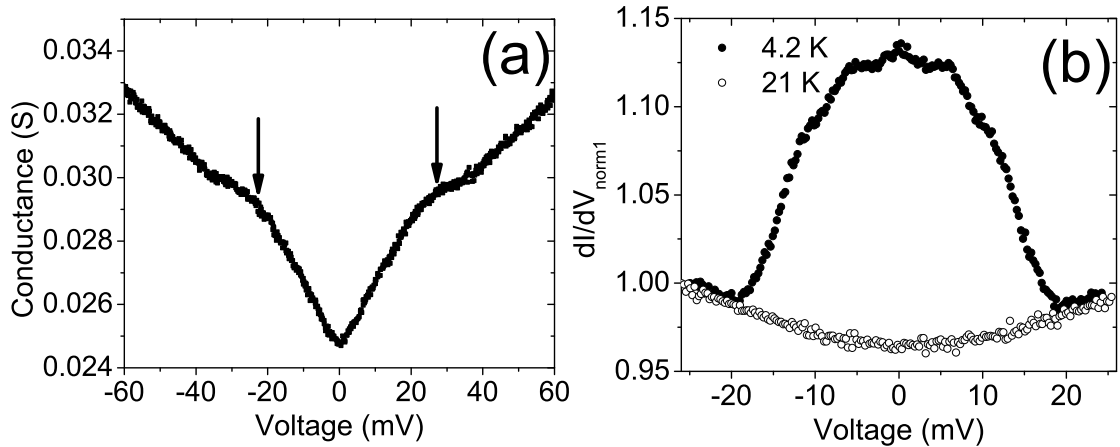}
  \caption{(a) $dI/dV-V$ curve for a high $Z$ measurement. (b) Normalized ($dI/dV_{norm1}-V$) point contact spectra for a low $Z$ junction at 4.2K and 21K. The two gaps at $\sim$6 meV and  $\sim$10 meV are marked with lines on the 4.2 K curve. The 21 K conductance curve shows the background conductance in the normal state.}
  \label{figure 2}
\end{figure}

The normalized $dI/dV_{norm2}-V$ curves were then fit to a two gap $s$-wave BTK model. Figure 3 shows the normalized data and the corresponding fits (solid lines) for temperatures between 4.2 K and 21 K.  The extracted gap sizes are 7.1 meV and 9.9 meV at 4.2 K, which is similar to the estimates from figure 2b i.e. before normalization with the 21 K data. Our fit has a $Z$ value of 0.05 for the small gap and 0.75 for the large gap, and $\Gamma$ values of 2.8 and 5.8 meV respectively at 4.2 K. Increasing the temperature causes the gap size to decrease linearly until 11 K when a sharp change in the slope occurs.  The gap size then continues to decrease in a more BCS-like manner until 19 K where we can no longer identify the gaps features.  This behavior is different from that observed by Tortello {\em et al.} ~\cite{PhysRevLett.105.237002}.  The inset of figure 3 shows the two gaps obtained from the fits versus temperature.

\begin{figure}
  \includegraphics[width=8.5cm]{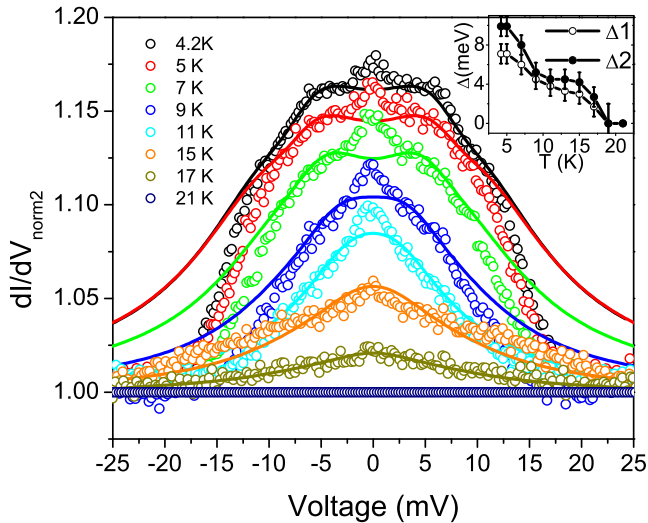}
  \caption{$dI/dV_{norm2}-V$ data at different temperatures from 4.2 K to 21 K. The extended BTK model fits are shown as solid lines.  The inset shows the variation of the two gaps as a function of temperature (Colour in online only).}
  \label{figure 3}
\end{figure}

Our normalized data show large flat top features which are consistent with an isotropic $s$-wave symmetry in a low $Z$ measurement, which had also been observed by Tortello {\em et al}~\cite{PhysRevLett.105.237002}.  To conclusively verify this symmetry structure we tested how the point contact spectra varied with a changing $Z$. A new point contact junction (different from the one used for Figs. 2 and 3) was formed and we confirm the reproducibility of the PCS data at low $Z$.
The $Z$ variation was measured on stable junctions starting with a low $Z$ then slowly retracting the tip causing an increase in $Z$. Due to the high risk of breaking a junction when the temperature is increased above $T_c$ we do not have a high temperature curve to normalize our data.  Instead we assume a V-shaped background with slopes matching the measured data at large voltages, this background is then thermally smeared as though it occurred at 21 K.  This smeared background is used to normalize the curves.  Shown in figure 4 are the normalized data with different $Z$: $Z$ = 0.05, 0.1, 0.14 for the small gap and $Z$ = 0.8, 0.9, 0.91 for the large gap (labeled as low $Z$, mid $Z$, and high $Z$ in figure 4).  From the fits, we consistently calculate a larger $Z$ value for the larger gap and a smaller $Z$ value for the smaller gap.  The behavior under increasing $Z$ is only consistent with an isotropic fully gapped system since the gap features show a broad flat behavior inside the gap (figure 1).  The extracted gaps in this measurement are 6.9 meV and 9.3 meV independent of our Z values.  We do find that as $Z$ increases we have an increasing $\Gamma$. We also note that while the low $Z$ measurements could only be fitted by a two gap system, the largest $Z$ data can be fitted with a single gap with magnitude between the magnitude of the two gaps and a massive increase in $\gamma$.  For this reason we found it preferable to extract the number of gaps from low $Z$ measurements and the symmetry of the gaps from higher $Z$ measurements.

\begin{figure}
  \includegraphics[width=8.5cm]{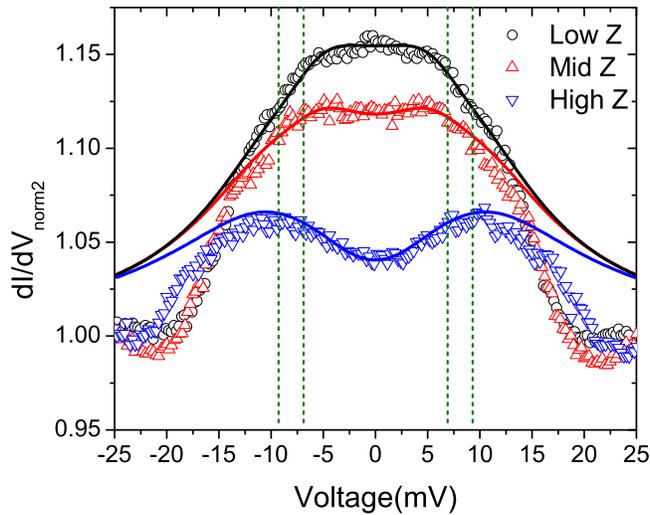}
  \caption{Normalized $dI/dV-V$ curves for different $Z$ values. The calculated $Z$ values are 0.05, 0.1, 0.14 (for the smaller gap) and 0.8, 0.9, 0.91 (for the larger gap) for the low $Z$, mid $Z$, and high $Z$ curves, respectively.  The gaps of 6.9 meV and 9.3 meV are marked by solid lines.(Colour in online only)}
  \label{figure 4}
\end{figure}

The small peak which appears at zero bias in the low $Z$ measurements (and can be seen in Figs. 2b, 3, and 4) should not be interpreted as a ZBCP resulting from a sign change in the order parameter since such a peak should increase in size with increasing $Z$.  In fact, the peak disappears as we increase $Z$ and we must look for another explanation for it.  One possible explanation is a Josephson peak resulting from tunnelling within the sample. To test whether the peak at zero bias is due to the Josephson effect we applied a magnetic field up to 5 T on a curve showing such a conductance peak.  The field did not remove the peak nor split it, only causing it to broaden very slightly.  It is also possible that it is caused by magnetic impurities, but as pointed out by Shan {\em et. al.} on a different pnictide~\cite{0295-5075-83-5-57004}, this would not be correlated with the superconducting transition and so the peak would not be suppressed upon reaching $T_c$.  One feasible explanation is that the peak is associated with pressure being applied by the tip on the sample.  This reason was given as a possible explanation for a similar peak seen in the 1111 compounds by Yates {\em et al.} ~\cite{0953-2048-21-9-092003}.  We find this to be a reasonable explanation for the peak observed in our measurements.

\section{Conclusions}

Our measurements show that BaFe$_{2-x}$Co$_{x}$As$_{2}$ ($x$ = 0.148) is a two gap superconductor.  The shapes of the point contact spectra as a function of $Z$ show that the gaps do not contain a sign flip as the characteristic ZBCP was not seen.  The existence of nodes is also unlikely since the spectra show features associated with full gaps.  We do not see evidence of nodes.  However, we cannot rule out the presence of a small amount of anisotropy.

\ack{Funding was provided by NSF DMR-0804452 (JT and AB) and DE-FG02-86ER45268 (JSK and GS).  We want to acknowledge Peter Hirschfeld for useful discussions and advice.}

\newpage

\noindent{\bf References}
\newline

\end{document}